# Spin-Orbit Torque in a Single Ferromagnetic Layer with Large Spin-Orbit Coupling

Ziyan Luo[1], Qi Zhang[1,2], Yanjun Xu[1], Yumeng Yang[1], Xinhai Zhang[2], and Yihong Wu[1,a)]

[1] *Department of Electrical and Computer Engineering, National University of Singapore, 4 Engineering Drive 3, Singapore 117583, Singapore*

[2] *Department of Electrical & Electronic Engineering, Southern University of Science and Technology, Xueyuan Rd 1088, Shenzhen 518055, China*

Spin-orbit torque in heavy metal/ferromagnet heterostructures with broken spatial inversion symmetry provides an efficient mechanism for manipulating magnetization using a charge current. Here, we report the presence of a spin torque in a single ferromagnetic layer in both asymmetric $MgO/Fe_{0.8}Mn_{0.2}$ and symmetric $MgO/Fe_{0.8}Mn_{0.2}/MgO$ structures, which manifests itself in the form of an effective field transverse to the charge current. The current to effective field conversion efficiency, which is characterized using both the nonlinear magnetoresistance and second-order planar Hall effect methods, is comparable to the efficiency in typical heavy metal/ferromagnet bilayers. We argue that the torque is caused by spin rotation in the vicinity of the surface via impurity scattering in the presence of a strong spin-orbit coupling. Instead of cancelling off with each other, the torques from the top and bottom surfaces simply add up, leading to a fairly large net torque, which is readily observed experimentally.

a) Electronic mail: elewuyh@nus.edu.sg

# I. INTRODUCTION

When a charge current passes through a ferromagnet (FM), due to the imbalance in electron density of states at the Fermi level and scattering asymmetry between spin-up and spin-down electrons, it becomes polarized, thereby generating a net spin current flowing in the charge current direction. In addition to longitudinal charge and spin currents, transverse charge and spin currents are also generated by the anomalous Hall effect (AHE), leading to charge and spin accumulation at the side surfaces or edges at steady-state [1]. So far, the study of AHE has been mainly focused on charge accumulation because it can be detected directly as a voltage signal, and very little attention was devoted to the spin accumulation. Recently, several groups have attempted to exploit the AHE-induced spin accumulation and related spin torque for magnetization switching applications in FM/nonmagnet (NM)/FM trilayers [2-5] which, compared to spin-orbit torque (SOT) generated by the spin Hall effect (SHE), offers the possibility of controlling the spin polarization direction by manipulating the magnetization direction of one of the FM layers. In addition, experiments have been carried out to detect the AHE-induced spin current through spin injection and nonlocal electrical detection [6-9]. We have recently demonstrated that, for a FM with large spin-orbit coupling (SOC), the spin accumulation would also result in an AHE-related magnetoresistance (MR), we termed it as anomalous Hall magnetoresistance (AHMR) [10]. The AHMR exhibits the same magnetization angle-dependence as that of spin Hall magnetoresistance (SMR) in heavy metal (HM)/FM [11-16], both of which show a linear dependence on current. In all these studies, the magnetization is typically in a saturation state, and therefore, nonlinear effect induced by spin torque, if any, is often suppressed by the large field and can hardly be observable.

The AHE occurs in solids either from extrinsic mechanisms like skew scattering and side-jump or intrinsic mechanism related to the Berry phase of electronic band structure [1, 17], with the extrinsic mechanism dominated in highly conductive ferromagnet. Regardless of its origin, it is commonly believed that spin polarization of the deflected electrons will follow the local magnetization direction (parallel for down-spin and antiparallel for up-spin) due to strong exchange interaction, so does the polarization of spins accumulated at the surfaces. Such kind of self-alignment of spin direction with the local magnetization facilitates the control of polarization direction of AHE-generated non-equilibrium spins using an external field, which is not possible for spin current generated by SHE. In the meantime, this also means that the accumulated spins near the surfaces are unable to exert any torque on the magnetization itself, limiting its practical applications. To circumvent this problem, several groups have proposed to employ a FM/NM/FM trilayer structure, in which the first FM serves as the spin generator whereas the second FM functions as a spin detector whose magnetization direction can be manipulated by the spin-transfer torque from the spin current generated by the first FM layer [2-5]. However, the spin current will decay while traveling across the NM layer, which works as a separator to decouple the two FM layers, resulting in a low efficiency. The same difficulty also faces FM/NM/FM trilayers in which spin current generated at the bottom FM layer and spacer layer interface exerts a torque on the top FM layer [18, 19]. Here, we demonstrate that a damping-like (DL) SOT exists in a single thin FM layer as long as it exhibits a sizable AHE arising from a large SOC. Specifically, for a thin $Fe_{0.8}Mn_{0.2}$ layer with in-plane anisotropy, we observed a DL SOT which manifests itself in the form of an effective in-plane field transverse to the charge current. The strength of the DL effective field is characterized by measuring both the nonlinear magnetoresistance and second-order planar Hall effect (PHE) induced by the SOT. For a symmetric structure of $MgO(2)/Fe_{0.8}Mn_{0.2}(5)/MgO(2)$, an effective field to current density ratio of 0.242 Oe/($10^{10}$A/m$^2$) is obtained, which is comparable to that of FL effective field in Pt/NiFe bilayers [20, 21]. We argue that part of the AHE-generated electron spins near the surfaces is misaligned with the magnetization direction due to precession upon scattering by the SOC scattering center, and the backflow of these misaligned spins exerts a torque on the magnetization. The effects from top and bottom surfaces simply add up rather than cancelling out with each other due to the scattering asymmetry of up-spin and down-spin electrons at the two surfaces, which eventually leads to an observable net spin torque.

## II. EXPERIMENTAL DETAILS AND RESULTS

### A. Nonlinear magnetoresistance measurement

Samples with the structure MgO($d_1$)/Fe$_{0.8}$Mn$_{0.2}$($d_2$)/sub and MgO($d_1$)/Fe$_{0.8}$Mn$_{0.2}$($d_2$)/MgO($d_1$)/sub are deposited on quartz substrate. The numbers inside the parentheses indicate thickness in *nm*. All the materials are prepared by magnetron sputtering with a base pressure of $2 \times 10^{-8}$ Torr and working pressure of $3 \times 10^{-3}$ Torr, respectively. Devices are patterned and contacted by Pt(10)/Cu(200)/Ta(5) (Ta is deposited first) electrodes using combined techniques of photolithography and lift-off. Both nonlinear MR and second-order PHE were measured to characterize the SOT. The MR measurements were performed using a Quantum Design Versalab Physical Property Measurement System. During the measurements, the sample is rotated relative to an in-plane field with fixed direction and then measure the nonlinear MR by applying an AC current along *x* direction. Instead of using the standard lock-in technique, we employed a Wheatstone full bridge to measure the nonlinear magnetoresistance [20, 22]. Compared to the lock-in technique, the latter is not affected by thermal drift and low-noise signals can be readily obtained without resorting to any post-measurement data processing. On the other hand, Hall bars are used for the second-order PHE measurements; more details will be presented when we discuss the PHE data.

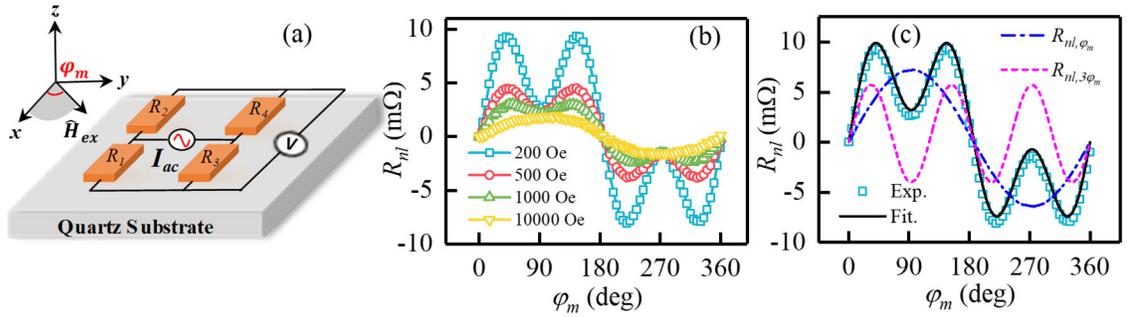

FIG. 1. (a) Schematic of Wheatstone bridge used for nonlinear MR measurement. (b) $R_{nl}$ vs $\varphi_m$ at different external fields. (c) Experimental (symbol) and fitting (solid-line) results for the $R_{nl}$ vs $\varphi_m$ curve obtained at $H_{ex} = 200$ Oe. Dash-dotted and dashed-lines refer to the decomposed $R_{nl,\varphi_m}$ and $R_{nl,3\varphi_m}$ components, respectively.

Figure 1(a) shows the schematic of a Wheatstone bridge which consists of four rectangular shaped elements with a dimension of 100 $\mu m$ (width) × 500 $\mu m$ (length). When an AC current, $I_{ac} = I_0 \sin \omega t$, is applied to two terminals of the bridge along $x$-axis, half of the current ($\frac{1}{2} I_0 \sin \omega t$) will flow in each element. Any nonlinear MR ($R_{nl}$) can be readily detected from the bridge output voltage which is given by:

$$V_{out} = \frac{1}{2} I_0 R_{nl} - \frac{1}{2} I_0 R_{nl} \cos 2\omega t. \qquad (1)$$

Here $R_{nl}$ is the change in resistance originated from both the SOT ($R_{nl}^{SOT}$) and thermoelectric effect ($R_{nl}^{\nabla T}$) [23, 24]. The effect of Oersted field can be excluded because it is almost cancelled out in a single metallic layer. As shown in Eq. (1), one can directly measure the DC component as the nonlinear signal because it is the same as the amplitude of the second harmonic signal, therefore, there is no need to use the lock-in technique [20, 22].

We first performed the angular dependent harmonic measurement at room temperature by applying a rotational field ($H_{ex}$) in the $xy$-plane with the strength varying from 100 Oe to 30 kOe. Figure 1(b) shows the typical result of $R_{nl}$ as a function of field angle ($\varphi_m$) of MgO(2)/Fe$_{0.8}$Mn$_{0.2}$(5). The shape of angle-dependence curve depends strongly on the strength of the external field, but in general can be fitted very well using the following equation:

$$R_{nl} = R_{nl,\varphi_m} \sin(\varphi_m - \varphi_0) + R_{nl,3\varphi_m} \sin 3(\varphi_m - \varphi_0) + R_0, \qquad (2)$$

where $R_0$ and $\varphi_0$ are the resistance and angle offsets due to experimental setup, and $R_{nl,\varphi_m}$ and $R_{nl,3\varphi_m}$ are the coefficient of $\sin \varphi_m$ and $\sin 3\varphi_m$ components, respectively. As an example, Fig. 1(c) compares the experimental and fitting results at $H_{ex} = 200$ Oe. The square symbols are the measured results and black solid-line is the overall fitting based on Eq. (2). The dash-dotted and dashed lines correspond to the 1$^{st}$ and 2nd terms of Eq. (2), respectively. The best fitting to Eq. (2) yields $R_{nl,\varphi_m} = 6.83$ mΩ, $R_{nl,3\varphi_m} = 4.91$ mΩ, $R_0$ = 3.14 mΩ, $\varphi_0 = 3.34°$. Based on previous studies of HM/FM bilayers, the symmetry of the signal shown in Fig. 1(b) suggests that there is an SOT effective field present in the single FM layer and its direction is along $y$-axis [23].

Without losing generality, the longitudinal resistance of each element can be considered as consisting of two parts: one is the linear resistance ($R_l$) which is independent of the current and the other is the nonlinear resistance ($R_{nl}$) which is proportional to the current. The former can be rewritten as:

$$R_l = R_z + (R_x - R_z)\sin^2\theta_m\cos^2\varphi_m + (R_y - R_z)\sin^2\theta_m\sin^2\varphi_m, \tag{3}$$

where $\theta_m$, $\varphi_m$ are the polar and azimuthal angles of $\hat{m}$, respectively, and $R_i$ is the longitudinal resistance measured when $\hat{m}$ is saturated along the direction $i = x, y, z$. The second term corresponds to the anisotropic magnetoresistance (AMR). As for the third term, although it exhibits a symmetry similar to that of spin Hall magnetoresistance (SMR) in HM/FM bilayer, as we demonstrated in the previous work, it has an AHE origin and is dubbed as anomalous Hall magnetoresistance (AHMR) [10]. On the other hand, the nonlinear part caused by the small changes in $\theta_m$ and $\varphi_m$ by $\Delta\theta_m$ and $\Delta\varphi_m$, due to current-induced effective field $\vec{H}_I$, is approximately given by

$$R_{nl}^{SOT} = (\Delta R_{xz}\cos^2\varphi_m + \Delta R_{yz}\sin^2\varphi_m)\sin 2\theta_m\Delta\theta_m + \Delta R_{yx}\sin^2\theta_m\sin 2\varphi_m\Delta\varphi_m, \tag{4}$$

where $\Delta R_{ij} = R_i - R_j(i, j = x, y, z)$. For a film with in-plane anisotropy, $\Delta\theta_m$ and $\Delta\varphi_m$ can be estimated using the first order approximation [23-27], i.e., $\Delta\theta_m \approx H_I^\theta / (H_{ex} + H_d)$ and $\Delta\varphi_m \approx H_I^\varphi / H_{ex}$; here, $H_I^\theta$ ($H_I^\varphi$) is the $\theta(\varphi)$ component of $\vec{H}_I$, $H_d$ is the out-of-plane demagnetizing field. Note that we have ignored the in-plane anisotropic field as it is much smaller than the external field. The contribution of $\Delta\theta_m$ induced change to $R_{nl}^{SOT}$, if any, should be very small as $\theta_m \approx \pi / 2$ and $H_I^\theta \ll H_d$. Therefore, we only need to consider the change in $R_{nl}^{SOT}$ caused by $\Delta\varphi_m$. When the SOT effective field is in negative *y*-direction (based on our measurement geometry), its projection in the azimuthal direction is given by $H_l^\varphi = -H_l \cos\varphi_m$. Since the effective field is proportional to current density, it can be written as $H_l = \alpha_{eff} j_{0\_rms}$; here, $\alpha_{eff}$ is the strength of effective field per unit current density in the FM layer, which represents the SOT generation efficiency, and $j_{0\_rms} = I_0/(2\sqrt{2}wd)$ is the root-mean-square (rms) amplitude of the applied AC current density, with *w* the width of individual elements and *d* the thickness of FM layer. Substitute all these parameters into Eq. (4), one obtains

$$R_{nl}^{SOT} = \frac{\Delta R_{xy}\sin^2\theta_m}{2H_{ex}}(\alpha_{eff}\sin\varphi_m)j_{0\_rms} + \frac{\Delta R_{xy}\sin^2\theta_m}{2H_{ex}}(\alpha_{eff}\sin 3\varphi_m)j_{0\_rms}. \qquad (5)$$

By adding to the thermoelectric contribution, $R_{nl}^{\nabla T} = \alpha\,\frac{l}{w}\frac{I_0}{2}\nabla T_z \sin\theta_m \sin\varphi_m$ [28], where $\alpha$ is the effective coefficient that accounts for anomalous Nernst effect (ANE) and spin Seebeck effect (SSE), *l* is the length of the individual elements, and $\nabla T_z$ is the thermal gradient along *z* direction, the total change of resistance induced by the current may be written as

$$R_{nl} = \frac{\Delta R_{xy}\sin^2\theta_m}{2H_{ex}}(\alpha_{eff}\sin\varphi_m)j_{0\_rms} + \frac{\Delta R_{xy}\sin^2\theta_m}{2H_{ex}}(\alpha_{eff}\sin 3\varphi_m)j_{0\_rms} + \alpha\,\frac{l}{w}\frac{I_0}{2}\nabla T_z \sin\theta_m \sin\varphi_m.$$

$$(6)$$

As mentioned above, when an AC current, $I_{ac} = I_0\sin\omega t$, is applied to two terminals of the bridge along *x*-axis, half of the current ($\frac{1}{2}I_0\sin\omega t$) will flow in each element. The Wheatstone bridge consists of 4 identical elements (by design), if we consider one of them first (*e.g.*, element #1), the voltage across its two terminal is given by

$$V_1 = [R_z + (R_x - R_z)\sin^2\theta_m\cos^2\varphi_m + (R_y - R_z)\sin^2\theta_m\sin^2\varphi_m$$
$$-(\Delta R_{yz} - \Delta R_{xz})\sin^2\theta_m\sin 2\varphi_m\,\frac{\alpha_{eff}j_{0\_rms}\sin\omega t\cos\varphi_m}{H_{ex}} + \alpha\,\frac{l}{w}\frac{I_0}{2}\sin\omega t\nabla T_z\sin\theta_m\sin\varphi_m]\frac{1}{2}I_o\sin\omega t.$$

$$\tag{7}$$

Similarly, for element #2, the voltage across its two terminals is given by (note the current is in opposite phase of element #1):

$$
\begin{aligned}
V_2 = &-[R_z + (R_x - R_z)\sin^2\theta_m \cos^2\varphi_m + (R_y - R_z)\sin^2\theta_m \sin^2\varphi_m \\
&+ (\Delta R_{yz} - \Delta R_{xz})\sin^2\theta_m \sin 2\varphi_m \frac{\alpha_{eff} j_{0\_rms} \sin\omega t \cos\varphi_m}{H_{ex}} - \alpha\frac{l}{w}\frac{I_0}{2}\sin\omega t \nabla T_z \sin\theta_m \sin\varphi_m]\frac{1}{2}I_o \sin\omega t
\end{aligned}
$$

$$\tag{8}$$

The algebraic sum of $V_1$ and $V_2$ gives the bridge output voltage

$$
\begin{aligned}
V_{out} = V_1 + V_2 &= [(\Delta R_{xz} - \Delta R_{yz})\sin^2\theta_m \sin 2\varphi_m \frac{\alpha_{eff} j_{0\_rms} \cos\varphi_m}{H_{ex}} + \alpha\frac{l}{w}\frac{I_0}{2}\nabla T_z \sin\theta_m \sin\varphi_m]I_o \sin^2\omega t \\
&= [(\Delta R_{xz} - \Delta R_{yz})\sin^2\theta_m \sin 2\varphi_m \frac{\alpha_{eff} j_{0\_rms} \cos\varphi_m}{2H_{ex}} + \alpha\frac{l}{w}\frac{I_0}{4}\nabla T_z \sin\theta_m \sin\varphi_m]I_o(1 - \cos 2\omega t)
\end{aligned}
$$

$$\tag{9}$$

From the time-average or DC output of the bridge voltage, we can obtain the nonlinear resistance:

$$
R_{nl} = \frac{\overline{V_{out}}}{\frac{1}{2}I_0} = R_{nl,\varphi_m}^{SOT}\sin\varphi_m + R_{nl,3\varphi_m}^{SOT}\sin 3\varphi_m + R_{nl,3\varphi_m}^{\nabla T}\sin\varphi_m. \tag{10}
$$

Here $R_{nl,\varphi_m}^{SOT} = R_{nl,3\varphi_m}^{SOT} = (\Delta R_{xy}/2H_{ex})\alpha_{eff} j_{0\_rms}$, and $R_{nl,\varphi_m}^{\nabla T} = \alpha\frac{l}{w}\frac{I_0}{2}\nabla T_z$. $R_{nl,\varphi_m}^{\nabla T}$ is independent of the external field as long as the external field is sufficiently large to saturate the magnetization in the field direction [23, 24, 28] (note: $\theta_m \approx \pi/2$). Eq. (10) reproduces the experimental results well except for the angle and resistance offsets which are caused by the sample fabrication and measurement processes rather than the intrinsic properties of the sample.

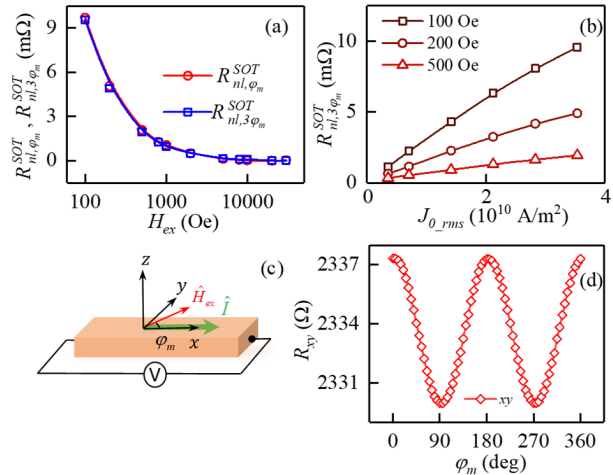

FIG. 2. (a) $R_{nl,\varphi_m}^{SOT}$ and $R_{nl,3\varphi_m}^{SOT}$ as a function of $H_{ex}$. Symbols are extracted from the decomposed fitting results. Solid-lines are the fitting results. (b) AC current density dependence of $R_{nl,3\varphi_m}^{SOT}$ at $H_{ex} = 100$, 200, and 500 Oe, respectively. (c) Geometry of angle-dependent magnetoresistance (ADMR) measurement in $xy$ plane. (d) Results of $R_{xy}$ as a function of $\varphi_m$ in $xy$ plane measured at $H_{ex} = 30$ kOe.

To examine the field dependence of $R_{nl,\varphi_m}^{SOT}$ and $R_{nl,3\varphi_m}^{SOT}$, we perform the angle-dependence fitting of all data measured at different external fields, and the extracted values for $R_{nl,\varphi_m}^{SOT}$ and $R_{nl,3\varphi_m}^{SOT}$ are plotted in Fig. 2(a) as a function of $H_{ex}$. As shown in the figure, $R_{nl,\varphi_m}^{SOT}$ and $R_{nl,3\varphi_m}^{SOT}$ decrease sharply as $H_{ex}$ increases in the low field range (< 5 kOe ), and at high filed, $R_{nl,\varphi_m}^{SOT}$ and $R_{nl,3\varphi_m}^{SOT}$ gradually diminishes. The solid-lines are the fitting results using $R_{nl,\varphi_m}^{SOT} = R_{nl,3\varphi_m}^{SOT} = (C_{SOT}/H_{ex})$ with $C_{SOT} = 970$ m$\Omega$·Oe (at $j_{0\_rms} = 0.71 \times 10^{10}$ A/m$^2$). The excellent agreement with Eq. (10) suggests that, indeed, there exists a SOT effective field in $y$-direction. To further ascertain its current related origin, we plot in Fig. 2(b) $R_{nl,3\varphi_m}^{SOT}$ as a function of current density at $H_{ex} = 100$, 200, and 500 Oe, respectively. As expected, an almost linear dependence is found between $R_{nl,3\varphi_m}^{SOT}$ and the current density at all field values. The slight deviation from the linear trend at high current density is presumably caused by the thermal effect. To estimate the SOT efficiency, we measured $\Delta R_{xy}$ of an individual element by aligning $\hat{m}$ with a rotating external field of 30 kOe in $xy$ plane as indicated in Fig. 2(c). The $R_{xy}$ as a function of $\varphi_m$ is plotted in Fig. 2(d). For a MgO(2)/Fe$_{0.8}$Mn$_{0.2}$(5) sample, $\Delta R_{xy}$ turned out to be 7.38 $\Omega$, from which $\alpha_{eff}$ is calculated as 0.372 Oe/($10^{10}$A/m$^2$); this value is comparable to the efficiency reported for Pt/NiFe bilayer at similar thickness range [20, 21].

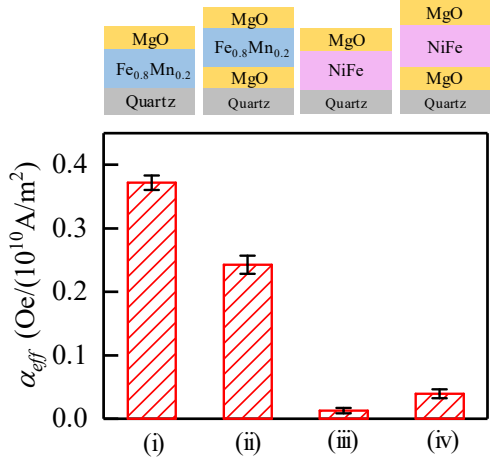

FIG. 3. Values of $\alpha_{eff}$ obtained for different structures: (i) MgO/Fe$_{0.8}$Mn$_{0.2}$, (ii) MgO/Fe$_{0.8}$Mn$_{0.2}$/MgO, (iii) MgO/NiFe, (iv) MgO/NiFe/MgO. The error bars indicate the fitting accuracy.

To shed light on the respective role of the FM and its capping layer and underlayer, we compare the $\alpha_{eff}$ values for the following four samples: i) MgO(2)/Fe$_{0.8}$Mn$_{0.2}$(5), ii) MgO(2)/Fe$_{0.8}$Mn$_{0.2}$(5)/MgO(2), iii) MgO(2)/NiFe(5), iv) MgO(2)/NiFe(5)/MgO(2). The rms amplitude and frequency of the applied AC current density were fixed at $0.71 \times 10^{10}$A/m$^2$ and 5000 Hz for all the samples. As shown in Fig. 3, the $\alpha_{eff}$ value for MgO(2)/Fe$_{0.8}$Mn$_{0.2}$(5)/MgO(2) is 0.242 Oe/($10^{10}$A/m$^2$); although it is smaller than that of MgO(2)/Fe$_{0.8}$Mn$_{0.2}$(5), the difference is not that large. Considering the fact that the crystalline structure and interface roughness for these two samples might be different, further structural analysis are required in order to reveal the true cause for this difference. In a sharp contrast, the $\alpha_{eff}$ values for MgO(2)/NiFe(5) is about one order of magnitude smaller than that of MgO(2)/Fe$_{0.8}$Mn$_{0.2}$(5). Again, there is no noticeable difference between the $\alpha_{eff}$ values for MgO(2)/NiFe(5) and MgO(2)/NiFe(5)/MgO(2) either. In fact, the $\alpha_{eff}$ value for MgO(2)/NiFe(5)/MgO(2) is even slightly larger than that of MgO(2)/NiFe(5). If we compare (i) with (iii) and (ii) with (iv), it is rather clear that, by simply changing the FM layer, the $\alpha_{eff}$ varies in a large magnitude. Therefore, $\alpha_{eff}$ depends more on the FM layer than the asymmetry of the interfaces.

These results suggest that Rashba-Edelstein effect can be excluded as the origin for the observed SOT [29-31], which if exists, would have been largely cancelled out in symmetric structures due to different sign of SOT at top/bottom surfaces.

## B. Second-order planar Hall effect measurement

The current-induced SOT indeed can induce nonlinear MR, but one cannot completely rule out other possibilities. To further test the relevance of SOT, we measure the current-induced effective field directly using the second-order planar Hall effect (PHE) technique [32-36], from which we extract $\alpha_{eff}$ and compare it with the values obtained from the nonlinear MR measurements. As illustrated in Fig. 4(a), a bias current $I$ is injected into a 200-$\mu m$-wide, 2000-$\mu m$-long Hall bar consisting of MgO(2)/Fe$_{0.8}$Mn$_{0.2}$(5)/MgO(2). A small transverse bias magnetic field $H_y$ generated by a pair of Helmholtz coils is superimposed with the effective field. The planar Hall voltage was measured at different $H_y$ with a sweeping external magnetic field $H_x$ in $x$-direction, and at a positive and a negative bias current, respectively. The second-order planar Hall voltage is calculated from the measured Hall voltages as

$$\Delta V_{xy}(H_y) = V_{xy}(+H_y, +I, H_{ex}) + V_{xy}(-H_y, -I, H_{ex}). \tag{11}$$

Under the small perturbation limit, $i.e.$, both the current induced field ($H_{eff}$) and applied transverse bias field ($H_y$) are much smaller than the external field ($H_{ex}$), the change in in-plane magnetization direction is proportional to $(H_y - H_{eff})/H_{ex}$. The linear dependence of second-order PHE voltage on the algebraic sum of $H_y$ and $H_{eff}$ allows one to determine the effective field by varying $H_y$ as both fields play an equivalent role in determining the magnetization direction. After some algebra, one can derive the relation $\Delta V_{xy}(H_y = 0)/[\Delta V_{xy}(-H_y) - \Delta V_{xy}(H_y)] = H_{eff}/2H_y$. Therefore, by linearly fitting $\Delta V_{xy}(H_y = 0)$ against $\Delta V_{xy}(-H_y) - \Delta V_{xy}(H_y)$, the ratio $H_{eff}/2H_y$ can be determined from the slope of the curve, which in turn can be used to extract $H_{eff}$. Figure 4(b) shows one set of second-order PHE voltage curves for MgO(2)/Fe$_{0.8}$Mn$_{0.2}$(5)/MgO(2) with bias current of 5mA at different $H_y$ (0, +0.5 and -0.5 Oe). The different magnitude of the signal is attributed to the change of the total field along $y$-direction. The ratio of $H_{eff}$ to $2H_y$ can be extracted from the slope of $\Delta V_{xy}(H_y=0)$ versus $\Delta V_{diff}$ plot, where $\Delta V_{diff} = \Delta V_{xy}(H_y = -0.5Oe) - \Delta V_{xy}(H_y = +0.5Oe)$, as shown in Fig. 4(c). The data for $H_{ex} < 30$ Oe are not included in the fitting in order to satisfy the requirement of small angle perturbation. Fig. 4(d) shows the dependence of $H_{eff}$ on current density. As expected, $H_{eff}$ changes linearly with the current density, and from the slope, the SOT efficiency, $\alpha_{eff}$, is extracted as $0.187$ Oe/($10^{10}$A/m$^2$); this is comparable to the value of $0.242$ Oe/($10^{10}$A/m$^2$) obtained from the nonlinear MR measurement.

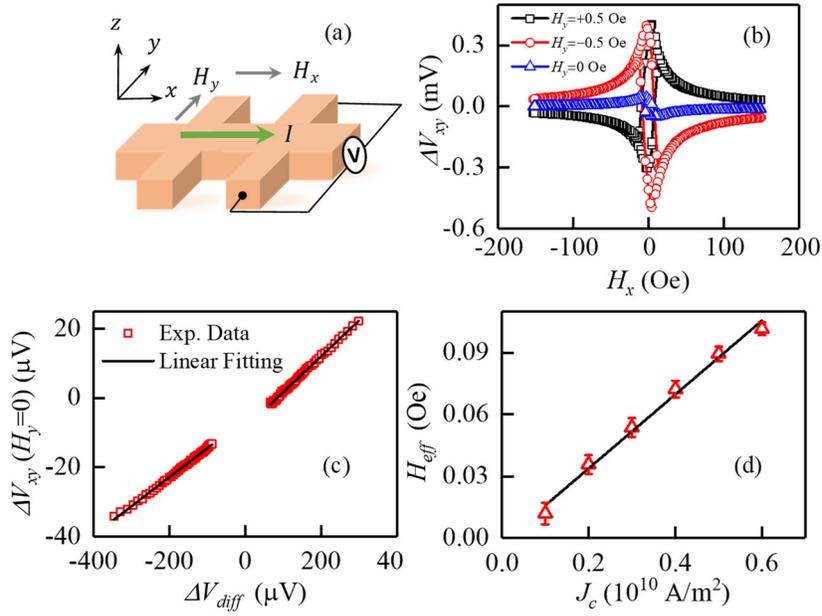

FIG. 4. (a) Schematic of the second-order PHE measurement. (b) Second-order planar Hall voltage at 5 mA bias current with different transverse fields $H_y$ ($0$, $+0.5$ and $-0.5$ Oe). (c) Linear fitting of $\Delta V_{xy}(H_y=0)$ against $\Delta V_{diff} = \Delta V_{xy}(H_y = -0.5Oe) - \Delta V_{xy}(H_y = +0.5Oe)$. (d) Linear dependence of $H_{eff}$ as a function of current density. Error bars indicate the accuracy of linear fitting.

## III. DISCUSSION

After the confirmation of presence of an effective field along $y$-axis by both techniques, we investigate how it depends on the Fe$_{0.8}$Mn$_{0.2}$ thickness by varying $d$ from 4 nm to 10 nm. As shown in Fig.

5(a), the results extracted from the two different techniques are fairly close to each other, with both scaling almost inversely with $d$, which implies that the SOT is of surface or interfacial nature. We have fabricated samples with even smaller thicknesses, but the sharp increase of resistance at small thickness does not allow to obtain reproducible results; therefore, we only focus on samples with a thickness larger than 4 nm. In our previous study on anomalous Hall effect induced magnetoresistance, *i.e.*, AHMR, we found that the absolute value of AHMR scales as $\left(\frac{\theta_{AH}}{\beta}\right)^2 \frac{2l_s}{d} \tanh\left(\frac{d}{2l_s}\right)$, where $d$ and $l_s$ are the thickness and spin diffusion length of FM, respectively, $\beta$ is the polarization for longitudinal conductivity, and $\theta_{AH}$ is the anomalous Hall angle [10]. If the SOT observed in this study is indeed from the backflow of reflected spins whose polarization is misaligned with the FM magnetization, the average SOT effective field should scale as $\frac{1}{d^2} \tanh\left(\frac{d}{2l_s}\right)$. The additional $1/d$ pre-factor comes from the fact that the torque generated is proportional to the total angular momentum transferred divided by the moment per unit area of the FM film. In Fig. 5(a), we plot the fitting results by using $l_s$ = 2-5 nm (shadowed region). The lower and upper boundary denoted the minimum and maximum values obtained by $l_s$ = 2 nm and $l_s$ = 5 nm. As can be seen, $l_s$ = 3.5 nm (green dotted line) fits the experimental results very well. This value is in good agreement with the value we obtained previously from AHMR studies.

Now the question is what could be the underlying mechanism for the observed SOT and related effective field? According to Manchon and Zhang [37], when a charge current flows through a single FM layer, the non-equilibrium transverse spin induced by the SOC exerts a torque on the local magnetization, but this is expected to be a small higher-order effect for inversion-symmetry-preserved SOC such as the impurity SOC and Luttinger spin-orbit band, which is presumably the present case. Recently, it has been theoretically predicted by Pauyac *et al.* [38] that transversely polarized spin current can exist in ferromagnet due to spin-orbit interaction, which generates spin current to compete with spin dephasing. Similarly, based on the first principles calculations, Amin *et al*. proposed that the intrinsic spin current is not subject to dephasing even though its spin polarization is misaligned with the magnetization [39], therefore, in addition to the AHE-induced spin current whose spin direction is along the magnetization direction, there is also an intrinsic spin current with the spin direction transverse to the magnetization in a single layer ferromagnet. Very recently, Wang *et al.* [40] experimentally observed the anomalous spin-orbit torques (ASOT) along *y*-direction in magnetic single-layer films (Py, Co, Ni and Fe), and attributed it to bulk SOI-generated transversely polarized spin current. Due to the exchange coupling in magnetic materials, the ASOT-induced magnetization tilting is observable when the FM is much thicker than the exchange length, whereas the ASOTs at top and bottom surfaces nearly cancel out in FM with thickness smaller than the exchange length. Although we don't completely rule out the relevance of bulk spin current and related torque, the thickness-dependence shown in Fig. 5(a) strongly suggests that that the SOT observed in this study is originated from the spin scattering and precession near the surface instead of bulk spin current. As discussed below, spin-dependent scattering and related spin rotation at the sample surface offers a more feasible explanation for the observed torque.

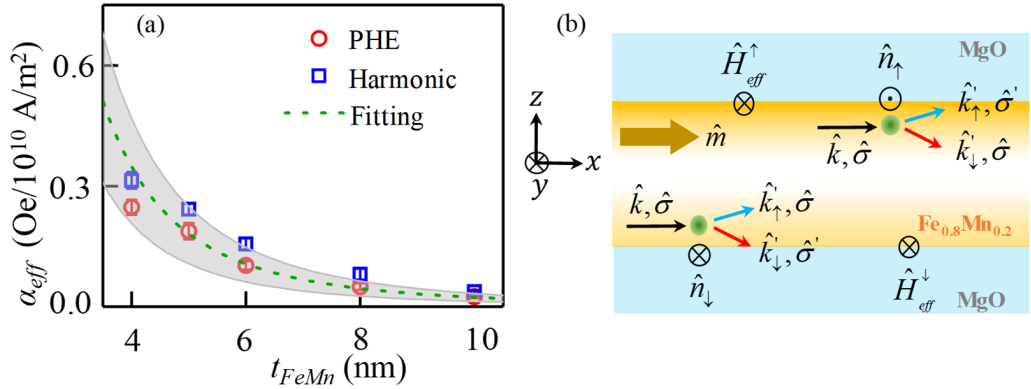

FIG. 5. (a) $Fe_{0.8}Mn_{0.2}$ thickness dependence of SOT efficiency $\alpha_{eff}$. Circles and squares are the values extracted from second-order PHE and nonlinear MR measurement, respectively. The shadowed area is the calculated $\alpha_{eff}$ range using $l_s$: 2-5 nm, and dotted-line is the fitting results using $l_s$ = 3.5 nm. (b) Schematic of effective fields generated at the top and bottom surface of $Fe_{0.8}Mn_{0.2}$ due to spin rotation.

As mentioned in the introduction, the polarization of AHE-generated spins inside the FM is presumably along the magnetization direction due to strong exchange coupling. However, in the vicinity of the interface, polarization direction of scattered electrons may deviate from the magnetization direction, depending on whether the electrons are scattered towards the surface or the interior of the FM layer; the latter will have its polarization aligned with the magnetization after it travels by a distance comparable to the exchange length, whereas the former will be reflected from the surface with its polarization determined by the scattering geometry and strength of the SOC of the scattering center, as illustrated in Fig. 5(b). When the current is in $x$-direction and considering the geometry of the sample, we may only need to focus on scattering in the $xz$ plane with its normal along $\hat{n} = \hat{k} \times \hat{k}'$ direction which is parallel to $y$-axis. Here, $\hat{k}$ ($\hat{k}'$) is electron moving direction before (after) scattering. For an electron with initial polarization direction $\hat{\sigma}$, upon scattering from an SOC scattering center, the polarization of scattered electrons may have the form $\hat{\sigma}' = S\hat{n} + T\hat{n} \times (\hat{\sigma} \times \hat{n}) + U(\hat{n} \times \hat{\sigma})$, where S, T and U are constants determined by the scattering angle and strength of SOC [41-44]. The spins polarized in the $\hat{\sigma}'$ direction will accumulate at the surface and then diffuse towards the film side, thereby exerting spin torques on the magnetization of the FM layer. Similar to the transfer of spin angular momentum from a nonlocal source to a local magnetic moment in HM/FM heterostructures, the spin torque here has the form of $\hat{\tau}_{ST} = \hat{m} \times (\hat{m} \times \hat{\sigma}')$ [45-48]. To facilitate discussion, we write $\hat{\sigma}$ as $\hat{\sigma} = sign(\sigma)\hat{m}$, where $sign(\sigma) = 1(-1)$ for down (up) spins. Then, the torque can be written as $\hat{\tau}_{ST} = [S - sign(\sigma)T\hat{m} \cdot \hat{n}]\hat{m} \times (\hat{m} \times \hat{n}) + sign(\sigma)U\hat{m} \times \hat{n}$. In the present case, since $\hat{m}$ is in the plane, the first terms is equivalent to an effective field along $z$-direction with its magnitude depending on the azimuthal angle of $\hat{m}$. This term can be ignored since the contributions from top and bottom surfaces have an opposite sign and thus are mostly cancelled out. On the other hand, the second term corresponds to a magnetization-independent effective field in $y$-direction with its direction depending on the direction of both $\hat{\sigma}$ and $\hat{n}$. We now consider the case that an electron moving in $x$-direction is scattered by an SOC scattering center near the top surface of the FM layer. As shown in the schematic of Fig. 5(b), we assume that up-spin electrons are scattered towards the surface side and down-spin electrons are scattered towards the film side, then $\hat{n}_\uparrow$ is in $-y$ direction for up-spin electrons and in $+y$ direction for down-spin electrons. This will lead to an effective field $\hat{H}_{eff}^\uparrow$ in $+y$ direction. If we assume that the scattering asymmetry remains the same for scattering near both surfaces, then the direction of $\hat{n}_\downarrow$ will be the same for both up-spin and down-spin electrons at the bottom surface. Since at the bottom surface, down-spin electrons will be scattered towards the surface side, the resultant SOT effective field $\hat{H}_{eff}^\downarrow$ is also in $+y$ direction. Although the torque from down-spin electrons may not be as large as that of up-spin electrons, when they add up it will lead to a sizable net torque. It is worth pointing out that based on this model, the field is of damping-like nature, but the effect on magnetization is equivalent to a field-like effective field along $y$ direction in HM/FM bilayers. And since the aforementioned process is similar to AHE in a toy model, the torque should be weak in samples with small AHE. In a previous work, we have demonstrated that the anomalous Hall angle of NiFe is around one order of magnitude smaller than that of $Fe_{0.71}Mn_{0.29}$ [10], which also corresponds to the size of anomalous Hall resistivity ($\rho_{xy}^{AH}$) in transition metals, *i.e.*, Fe >> Ni [49, 50]. This explains why the SOT is weak in NiFe control samples. The spin rotation is expected to be of surface nature and is present only within the spin dephasing length. More detailed theoretical studies are required to quantify the spin accumulation caused by the spin rotation as well as the resultant torque.

## IV. CONCLUSIONS

In summary, spin torque with a strength comparable to the field-like SOT in Pt/NiFe bilayers has been observed in both asymmetric $MgO/Fe_{0.8}Mn_{0.2}$ and symmetric $MgO/Fe_{0.8}Mn_{0.2}/MgO$ structures. The thickness dependence of current-to-torque conversion efficiency suggests that the torque is originated from surface or interfaces with the oxides. Although we don't rule out other possibilities, we argue that the torque is caused by spin rotation in the vicinity of the surface via impurity scattering in the presence of a strong spin-orbit coupling. Instead of cancelling off with each other, the torques from the top and bottom surfaces simply add up, leading to the observed torque. We believe our work will stimulate more studies on spin current and spin torque in single layer ferromagnet with large spin orbit coupling.

## ACKNOWLEDGMENTS

The authors would like to acknowledge support by Ministry of Education, Singapore under its Tier 2 Grants (grant no. MOE2017-T2-2-011 and MOE2018-T2-1-076), and stimulating discussions with Shufeng Zhang of The University of Arizona.